\documentclass[procedia]{easychair}
\usepackage{doc}
\usepackage{makeidx}

\usepackage{amsmath}% \begin{align}\end{align}
\usepackage{amssymb}% 
\usepackage{graphicx}% Include figure files
\usepackage{bm}% bold math
\usepackage{url}
\usepackage{ulem}
\usepackage{hyperref}% add hypertext capabilities

\def\procediaConference{99th Conference on Topics of Superb Significance (COOL 2014)}

\newcommand{\lag}{\langle}
\newcommand{\rag}{\rangle}
\newcommand{\lt}{\left}
\newcommand{\rt}{\right}

\newcommand{\hspc}{\hspace{1em}}
\newcommand{\htab}{\hspace{2em}}

\newcommand{\mrm}{\mathrm}
\newcommand{\mcl}{\mathcal}

\title{Quasi-Spin Correlations in a Frustrated Quantum\\ Spin Ladder}

\author{
    Takanori Sugimoto\inst{1}%\thanks{Designed and implemented the class style}
\and
    Michiyasu Mori\inst{2}%\thanks{Did numerous tests and provided a lot of suggestions}
\and
    Takami Tohyama\inst{1}%\thanks{Masterminded EasyChair and created versions 3.0--3.4 of the class style}
\and \\
    Sadamichi Maekawa\inst{2,3}
}

\institute{
  Tokyo University of Science,
  Katsushika, Tokyo, Japan\\
\and
   Japan Atomic Energy Agency,
   Tokai, Ibaraki, Japan\\
   \and
   ERATO, Japan Science and Technology Agency, Sendai 980-8577, Japan
 }
\authorrunning{Sugimoto, Mori, Tohyama and Maekawa}

\begin{document}

\maketitle


\begin{abstract}
The quasi-spin correlations in a frustrated quantum spin ladder with one-half magnetization are theoretically studied by using the density-matrix renormalization-group method and the quasi-spin transformation.
In this model, the frustration induces a gapless-to-gapful phase transition with a strong rung coupling. 
The gapful state is observed as the one-half magnetization plateau in the magnetization curve.
In the magnetization-plateau state, we find that the quasi-spin dimer has a large expectation value with long-ranged correlations.
This result does not only comes in useful to clarify the magnetization-plateau state, but gives a crucial information to understand the magnetization curve of the real compound BiCu$_2$PO$_6$, whose effective spin model corresponds to ours.  
\end{abstract}

%\setcounter{tocdepth}{2}
%{\small
%\tableofcontents}

%\section{To mention}
%
%Processing in EasyChair - number of pages.
%
%Examples of how EasyChair processes papers. Caveats (replacement of EC
%class, errors).

%------------------------------------------------------------------------------
\section{Introduction}
\label{sec:intro}
Quantum phase transitions and correlation functions in quantum spin systems have been attracted much attention due to their potential variety of quantum phenomena.
For instance, previous works have shown an emergence of magnon-based Bose-Einstein condensation near a saturation field~\cite{nikuni00,ruegg03,giamarchi99,giamarchi08,zapf14}, and spinon-based metal-insulator transitions induced by a magnetic field~\cite{oshikawa97}.
In these cases, the quantum spins behave like either a bosonic object (magnon) or a fermionic one (spinon), where the magnetic field plays a role of a chemical potential for the magnons or spinons.
In addition, the present authors have recently presented another type of spin character in a magnetic field, where the quantum spins are reconstructed into quasi-spins with an effective magnetic field~\cite{sugimoto15}.
The reconstruction reproduces a quasi-spin Hamitonian that is very similar to the original Hamitonian.
Therefore, several resemblaces appear in physical quantities with a revision of the quasi-spins as magnetic objects.

Recently, a frustarted quantum spin ladder (F-QSL) has been investigated, because it is expected to correspond to an effective spin model of the real compound BiCu$_2$PO$_6$~\cite{tsirlin10,lavarelo12,sugimoto13} and several magnetic phase transitions have been observed in the compound~\cite{sugimoto15,kohama12,casola13,kohama14,sugimoto14}.
Since some of the magnetic phase transitions cannot be understood without a frustration, an effect of the frustration should be crucial to clarify the phase transitions.
In practice, with a strong frustration, the present authors have found some magnetic phase transitions with magnetization plateaux in the same model.
In order to understand the magnetization curve of BiCu$_2$PO$_6$, it is important to examine their magnetic phases in detail one by one.
In this paper, we study a plateau of magnetization $M/M_{\mrm{sat}}=1/2$, where $M_{\mrm{sat}}$ denotes the saturated magnetization. 

%------------------------------------------------------------------------------
\section{Effective Hamiltonian in the strong rung-coupling limit}

A Hamiltonian of the F-QSL normalized by the rung-coupling $J$ in a magnetic field $h$ is given by,
\begin{equation}
\mcl{H}/J=\mcl{H}_\parallel+\mcl{H}_\perp+\mcl{H}_{Z}, \label{eq:ham1}
\end{equation}
with
\begin{align}
\mcl{H}_\parallel&=\lambda \sum_{i=\mrm{u,l}}\lt(\sum_{j=0}^{L-2}\bm{S}_{j,i}\cdot\bm{S}_{j+1,i} +\sum_{j=0}^{L-3}\eta\, \bm{S}_{j,i}\cdot\bm{S}_{j+2,i}\rt),\\
\mcl{H}_\perp&=\sum_{j=0}^{L-1}\bm{S}_{j,\mrm{u}}\cdot\bm{S}_{j,\mrm{l}}, \\
\mcl{H}_{Z}&= -h \sum_{j=0}^{L-1} \sum_{i=\mrm{u,l}} {S}_{j,i}^z,
\end{align}
where $\bm{S}_{j,\mrm{u(l)}}$ ($j=0,1,\cdots,L-1$) is the $S=1/2$ spin operator on the $j$ site in the upper (lower) chain and its $z$ component is $S^z_{j,\mrm{u(l)}}$.
Here, $\lambda$ denotes a nearest neighbor magnetic exchange coupling in the leg normalized by $J$, and the second-neighbor one is denoted by $\eta$, which plays a role of the geometrical frustration in the leg.
The normalized magnetic field is positively defined as $h>0$.
Below, we consider only the anti-ferromagnetic interactions, i.e., $\lambda>0$, $\eta>0$, and the strong rung-coupling limit, $\lambda \ll 1$.

\subsection{Quasi-spin transformation}
In the strong rung coupling limit $\lambda \ll 1 $ with finite magnetizations, a quasi-spin transformation and a reduced Hamiltonian are useful to understand the ground state and the low-energy physics.
In order to obtain the reduced Hamiltonian, we start with two spin Hamiltonian on a rung as follows,
\begin{equation}
\mcl{H}_{\mrm{rung}}=\bm{S}_{\mrm{u}}\cdot\bm{S}_{\mrm{l}}-h\sum_{i=\mrm{u,l}}\bm{S}_{i}=\frac{1}{2} (d_{\mrm{u}}^\dagger d_{\mrm{l}} + d_{\mrm{l}}^\dagger d_{\mrm{u}}) +\lt(n_{\mrm{u}}-\frac{1}{2} -h\rt)\lt(n_{\mrm{l}}-\frac{1}{2} -h\rt)+h^2,
\end{equation}
with a Jordan-Wigner transformation of spin operators,
\begin{equation}
d_{\mrm{u}} = S_{\mrm{u}}^- e^{- i\frac{\pi}{2}(S_{\mrm{l}}+\frac{1}{2})}, \hspc d_{\mrm{l}} = S_{\mrm{l}}^- e^{+ i\frac{\pi}{2}(S_{\mrm{u}}+\frac{1}{2})},
\end{equation}
and the number operators $n_{\mrm{u}(\mrm{l})}=d_{\mrm{u}(\mrm{l})}^\dagger d_{\mrm{u}(\mrm{l})} = S_{\mrm{u}(\mrm{l})}^z+\frac{1}{2}$.
To diagonalize this Hamiltonian, we can use a bonding and an anti-bonding operators for create and annihilate operators of Jordan-Wigner fermions, $d_{\mrm{b}} = (d_{\mrm{u}} + d_{\mrm{l}})/\sqrt{2}$, $d_{\mrm{a}} = (d_{\mrm{u}} - d_{\mrm{l}})/\sqrt{2}$. 
With the number operator of the bonding and an anti-bonding Jordan-Wigner fermions $n_{\mrm{b}(\mrm{a})}$, the rung Hamiltonian is rewritten as,
\begin{equation}
\mcl{H}_{\mrm{rung}} = n_{\mrm{a}}(n_{\mrm{b}}-1)-h(n_{\mrm{a}}+n_{\mrm{b}}-1) +\frac{1}{4}.
\end{equation}
We can obtain a quasi-spin transformation with an inverse Jordan-Wigner transformation as follows,
\begin{equation}
T_{\mrm{p}}^+ = id_{\mrm{b}}^\dagger, \hspc T_{\mrm{p}}^- = -i d_{\mrm{b}}, \hspc T_{\mrm{p}}^z= n_{\mrm{b}}-\frac{1}{2},
\end{equation}
and
\begin{equation}
T_{\mrm{m}}^- = d_{\mrm{a}}^\dagger e^{-i\pi n_{\mrm{b}}}, \hspc T_{\mrm{m}}^+ = d_{\mrm{a}} e^{i\pi n_{\mrm{b}}}, \hspc T_{\mrm{m}}^z = \frac{1}{2} - n_{\mrm{a}}.
\end{equation}
These operators also satisfy the spin SU(2) algebra for themselves, and commutate each other.
The quasi-spin operators rewrite the rung Hamiltonian as,
\begin{equation}
\mcl{H}_{\mrm{rung}}= -\lt(T_{\mrm{p}}^z-\frac{1}{2}\rt)\lt(T_{\mrm{m}}^z-\frac{1}{2}\rt)-h(T_{\mrm{p}}^z-T_{\mrm{m}}^z) +\frac{1}{4} = -h_{\mrm{p}} T_{\mrm{p}}^z +  h_{\mrm{m}} T_{\mrm{m}}^z,
\end{equation}
where effective quasi-magnetic fields,
\begin{equation}
h_{\mrm{p}} \equiv h -\frac{1}{2}, \hspc h_{\mrm{m}} \equiv h +\frac{1}{2} - T_{\mrm{p}}^z > h.
\end{equation}
With the quasi-spin operators, the leg Hamiltonian between $j$-th and $k$-th rungs is given by,
\begin{align}
\mcl{H}_{\mrm{leg}} = \sum_{i=\mrm{u,l}} \bm{S}_{j,i}\cdot\bm{S}_{k,i} &= \frac{1}{2} \lt(T_{j,\mrm{p}}^z-T_{j,\mrm{m}}^z\rt)\lt(T_{k,\mrm{p}}^z-T_{k,\mrm{m}}^z\rt)-\lt(T_{j,\mrm{p}}^+ T_{j,\mrm{m}}^+-T_{j,\mrm{p}}^- T_{j,\mrm{m}}^-\rt)\lt(T_{k,\mrm{p}}^+ T_{k,\mrm{m}}^+-T_{k,\mrm{p}}^- T_{k,\mrm{m}}^-\rt) \notag\\
&\htab +\frac{1}{2}\Bigg\{T_{j,\mrm{p}}^+ T_{k,\mrm{p}}^-\cos\lt[ \frac{\pi}{2}(T_{j,\mrm{m}}^z-T_{k,\mrm{m}}^z)\rt] +T_{j,\mrm{m}}^+ T_{k,\mrm{m}}^-\cos\lt[ \frac{\pi}{2}(T_{j,\mrm{p}}^z-T_{k,\mrm{p}}^z)\rt] \notag\\
&\htab\htab - T_{j,\mrm{p}}^+ T_{k,\mrm{m}}^+ \cos\lt[ \frac{\pi}{2}(T_{j,\mrm{m}}^z-T_{k,\mrm{p}}^z)\rt] -T_{j,\mrm{m}}^- T_{k,\mrm{p}}^-\cos\lt[ \frac{\pi}{2}(T_{j,\mrm{p}}^z-T_{k,\mrm{m}}^z)\rt] +\mrm{H. c.}  \Bigg\}.
\end{align}
If we consider small quasi-magnetic field for $T_{\mrm{p}}$ spin, namely $|h_{\mrm{p}}| \ll 1$, the quasi-magnetic field for $T_{\mrm{m}}$ spin is much larger than $|h_{\mrm{p}}|$, namely $|h_{\mrm{m}}|\gtrsim \frac{1}{2}\gg |h_{\mrm{p}}|$.
In order to deal low-energy physics, we can project out the high-energy states, that is, quasi-up-spins of $T_{\mrm{m}}$ opeartors.
With the projection operator given by $\mcl{P}=\prod_j \lt(T_{j,\mrm{m}}^z-\frac{1}{2}\rt)$, an effective Hamiltonian is obtained as $\mcl{P}\mcl{H}\mcl{P}=\mcl{H}_{\mrm{eff}}$.
Since the original Hamiltonian (\ref{eq:ham1}) is composed by a sum over the rung and leg Hamiltonians, $\mcl{H}_{\mrm{rung}}$ and $\mcl{H}_{\mrm{leg}}$, the quasi-spin transformation gives us the effective Hamiltonian as follows,
\begin{equation}
\mcl{H}_{\mrm{eff}}/J^\prime = \mcl{H}_\parallel^\prime  + \mcl{H}_Z^\prime, \label{eq:ham2}
\end{equation}
with
\begin{align}
\mcl{H}_\parallel^\prime &= \lt[ \sum_{j=0}^{L-2} \lt(\bm{T}_{j}, \bm{T}_{j+1}\rt)_\Delta +\sum_{j=0}^{L-3}\eta \,\lt(\bm{T}_{j}, \bm{T}_{j+2}\rt)_\Delta \rt], \\
\mcl{H}_Z^\prime &= - h^\prime \sum_{j=0}^{L-1} T_{j}^z,
\end{align}
where $\bm{T}_{j}$ is an abbreviation of $\bm{T}_{j,\mrm{p}}$, and the effective magnetic field is obtained as $h^\prime = h-1-\frac{\lambda}{2}(1+\eta)$. Here, anisotropic exchange interactions $\lt(\bm{T}_{j}, \bm{T}_{k}\rt)_\Delta$ are defined as,
\begin{equation}
  \lt(\bm{T}_{j}, \bm{T}_{k}\rt)_\Delta \equiv \Delta T_j^zT_k^z + \frac{1}{2} (T_j^+T_k^- + T_j^-T_k^+), 
\end{equation}
where the anisotropy of the quasi-spin exchange interactions is given by $\Delta=1/2$.
For the aim of calculations, we note that the explicit forms of the quasi-spin operators on $j$-th rung $\bm{T}_j$ is given by,
\begin{align}
T_{j}^\pm&=\pm\frac{i}{\sqrt{2}}\left[S_{j,\mrm{u}}^\pm e^{\pm i\frac{\pi}{2}\left(S_{j,\mrm{l}}^z+\frac{1}{2}\right)}+S_{j,\mrm{l}}^\pm e^{\mp i\frac{\pi}{2}\left(S_{j,\mrm{u}}^z+\frac{1}{2}\right)}\right],\label{tr1}\\
T_{j}^z&=\frac{1}{2}\left[(S_{j,\mrm{u}}^z+S_{j,\mrm{l}}^z)+S_{j,\mrm{u}}^+S_{j,\mrm{l}}^-+S_{j,\mrm{u}}^-S_{j,\mrm{l}}^+\right],\label{tr2}
\end{align}
Therefore, our problem of the frustrated spin ladder (\ref{eq:ham1}) is reduced to the anisotropic spin chain (\ref{eq:ham2}) using the transformations (\ref{tr1}) and (\ref{tr2}). 
This fact makes it easier to analyze some correlation functions. 

\subsection{Quasi-spin correlations}
\label{subsec:corr}
In this paper, we examine two types of correlation functions of quasi-spin operators, namely, spin-spin and dimer-dimer correlation functions.
These correlation functions will show us a phase transition between a gapless spin-liquid phase and a gapful dimer phase of the frustrated spin chain system.
Since the frustrated spin chain system is the effective Hamiltonian of the F-QSL in the strong rung-coupling limit with $M/M_{\rm sat}=$1/2, those correlation functions of quasi-spins will clarify the gapless-to-gapful phase transition of the F-QSL.

We define a correlation functions between two quasi-spin operators,
\begin{equation}
C_{\mrm{t}}^z(r_j)=\lag T_{j_L}^z T_{j_R}^z\rag,
\end{equation}
where $j_L$ ($j_R$) denotes the site-index counted from the left (right) end of the chain, and  determines the distance between two sites $r_j=1,2,\cdots,L-1$,
\begin{equation}
j_L=\frac{L}{2}-2-\mrm{int}\lt(\frac{r_j}{2}\rt), \hspc j_R=\frac{L}{2}-2+\mrm{int}\lt(\frac{r_j+1}{2}\rt),
\end{equation}
where $L$ is the number of rungs.

The dimer-dimer correlation functions of quasi-spin operators is given by,
\begin{equation}
C_{\mrm{td}}^z(r_j)=\lag D_{j_L}^z D_{j_R}^z\rag,
\end{equation}
where $D_{j}^z$ ($j=1,2,\cdots,L-2$) is a $z$ component of a quasi-spin dimer operator defined as,
\begin{equation}
D_j^z = T_{j-1}^z T_{j}^z - T_{j}^z T_{j+1}^z.
\end{equation}

Below, we use the following definition of the expectation values of the quasi-spin and quasi-dimer operators in the real space as, $E_{\mrm{t}}^z(x_j) = \lag T_{j}^z\rag$ and $E_{\mrm{td}}^z(x_j) = \lag D_{j}^z\rag.$

%------------------------------------------------------------------------------
\section{Numerical results}
\label{sec:result}
\subsection{Bound energy}
\begin{figure}
\centering
\includegraphics[width=1.0\textwidth]{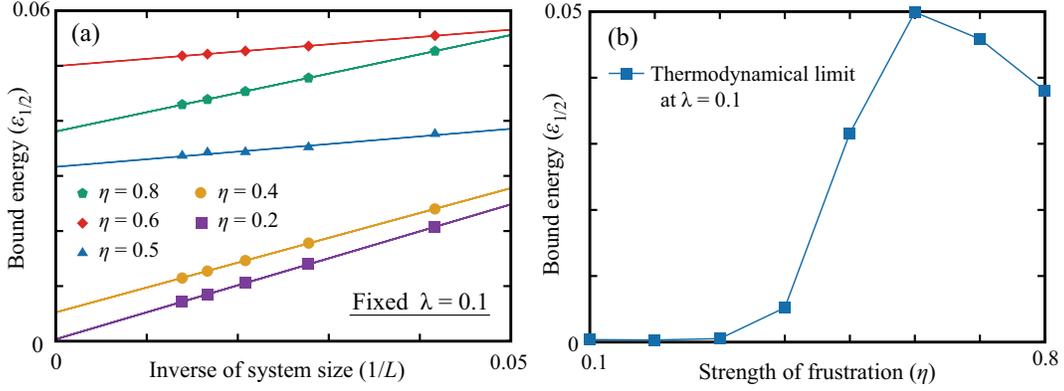}
\caption{(Color online) (a) Bound energy versus inverse of system size for various frustrations and (b) frustration dependence of the bound energy in the thermodynamical limit $L\to\infty $. In Fig.~(a), solid lines are determined by the least squares method.}
\label{fig1}
\end{figure}

In order to obtain the critical point of the gapless-to-gapful phase transition in the strong rung-coupling limit, we calculate the bound energies at $M/M_{\mrm{sat}}=1/2$ with a fixed $\lambda=0.1$ for various frustrations by using the density-matrix renormalization-group (DMRG) method (See Fig.~\ref{fig1}).
The bound energy at $M=M_{\mrm{sat}}/2$ is defined as,
\begin{equation}
\varepsilon_{1/2} = (E_{M_{\mrm{sat}}/2+1}+E_{M_{\mrm{sat}}/2-1}-2E_{M_{\mrm{sat}}/2})/J,
\end{equation}
where $E_{M}$ is the lowest energy with a magnetization $M$.
If there is a magnetization plateau at $M/M_{\mrm{sat}}$, the bound energy $\varepsilon_{M/M_{\mrm{sat}}}$ corresponds to the width of the flat region in the magnetization curve.
In Fig.~\ref{fig1} (a), the bound energy with a weak frustration $\eta=0.1$ is extrapolated to zero, although a strong frustration $\eta\geq 0.4$ induces a finite bound energy in the thermodynamical limit. 
We find a critical point of the gapless-to-gapful phase transition between $\eta=0.3$ and $0.4$ in Fig.~\ref{fig1} (b).　
The bound energy turns to decrease as the frustration increases over $\eta\cong 0.6$, because the ground state should approach another gapless phase in decoupled quasi-spin chains in the strong frustration limit $\eta\to \infty$. 

\subsection{Correlation functions}
We also calculate the expectation values and correlation functions of quasi-spin and quasi-dimer operators by using the DMRG method. 
% (Fig.~\ref{fig1}).
These are calculated in a 72-rung F-QSL under the open boundary condition for two values of the frustration $\eta=0.1$ and $0.6$, with a fixed $\lambda=0.1$.
Figure 2 (a) shows that real-space distributions of the expectation values of quasi-spin and quasi-dimer operators.
In this figure, we can see that the quasi-spins localize around the edges of the system and the expectation values ($E_{\mrm{t}}^z$) monotonically decrease to the center of the system for both $\eta=0.1$ and $0.6$.
On the other hand, the quasi-dimer operator has a large expectation value for $\eta=0.6$ as compared with $\eta=0.1$.
This character is also confirmed by the correlation functions in Fig.~2~(b).
We can see that the correlation functions of quasi-spin and dimer operators decay in the power law with a weak frustration $\eta=0.1$.
Thus, the $M/M_{\mrm{sat}}=1/2$ state with $\eta=0.1$ is understood as a gapless spin-liquid state of the quasi-spins, where the any correlation functions of the quasi-spins decays in the power low. 
With a strong frustration $\eta=0.6$, however, the correlation function of the quasi-dimer opeartor ($C_{\mrm{td}}^z$) retains a finite value in the infinite length limit, although the correlation function of quasi-spin operator ($C_{\mrm{t}}^z$) rapidly decays.
These behaviors indicate that the Majumdar-Gosh state~\cite{majumdar69} of the quasi-spins emerges at $\eta=0.6$, where an energy gap appears between the magnetized states $M=M_{\mrm{sat}}/2$ and $M=M_{\mrm{sat}}/2\pm 1$.
In this state, there is a valence-bond solid of dimers constructed by a singlet and a triplet on neighboring rungs.

\begin{figure}
\centering
\includegraphics[width=1.0\textwidth]{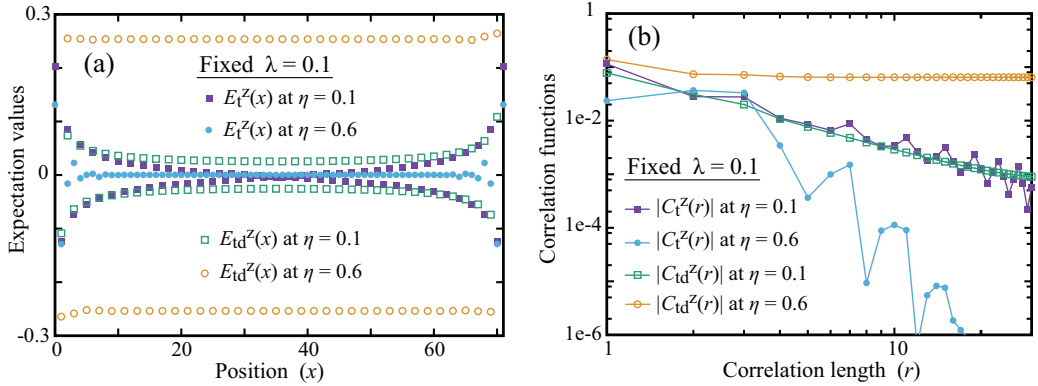}
\caption{(Color online) (a) Expectation values and (b) absolute values of correlation functions of quasi-spins for $\eta=0.1$ and $0.6$, with a fixed $\lambda=0.1$ in a 72-rung F-QSL.  The magnetization of the system is fixed at $M/M_{\mrm{sat}}=1/2$, and the truncation number of DMRG calculation is set on $m=300$.}
\label{fig2}
\end{figure}

%------------------------------------------------------------------------------
\section{Summary and discussions}
\label{sec:result}
We theoretically investigate the quasi-spin correlations in the 1/2 magnetized states of the frustrated quantum spin ladder with a strong rung coupling.
The effective model of these states corresponds to a frustrated spin chain with an Ising anisotropy.
With a weak frustration, the ground state has no gap, and is understood as a spin-liquid state, where the quasi-spin correlations decay in the power low.
On the other hand, a strong frustration changes the ground state into a gapful one, where the dimer-dimer correlation of the quasi-spins retains a finite value in the infinite length limit, although the quasi-spin correlation rapidly decays.
Since these behaviors correspond to those of the real spins in the frustrated spin chain, we conclude that the quasi-spins in the effective model gives a valence-bond-solid picture which helps us to understand the 1/2 magnetized states with a strong frustration.
This picture is not only useful to determine the critical points of the other gapless-to-gapful phase transition, but also will give a hint to understand the magnetization curve of the real compound BiCu$_2$PO$_6$, whose effective spin model corresponds to ours.

%------------------------------------------------------------------------------
\subsection*{Acknowledgments}
\label{sec:acks}
We would like to thank M. Fujita and O. P. Sushkov for valuable discussions. 
This work was partly supported by the Grant-in-Aid for Scientific Research 
(No.25287094, No.26108716, No.26247063, No.26103006, No.26287079, No.15K05192, and No.15H03553) from JSPS and MEXT, the HPCI Strategic Programs for Innovative Research (SPIRE) (hp140215) and the Computational Materials Science Initiative (CMSI), and by the inter-university cooperative research program of IMR Tohoku University.
Numerical computation in this work was carried out on the supercomputers at JAEA, the K computer at the RIKEN Advanced Institute for Computational Science, and the Supercomputer Center at Institute for Solid State Physics, University of Tokyo.

%------------------------------------------------------------------------------
% Refs:
%
\label{sec:bib}
\bibliographystyle{plain}
\end{document}